\documentclass[aps,prb,superscriptaddress,showpacs,preprint]{revtex4-1}

\usepackage{graphicx}
\usepackage{color}
\usepackage{hyperref}
\usepackage{pslatex}
\usepackage{amsmath}
\usepackage{longtable}

\begin{document}

\title{Highly anisotropic interlayer magnetoresitance in ZrSiS nodal-line Dirac semimetal    }
\author{M. Novak}
\email{mnovak@phy.hr}
\affiliation{Department of Physics, Faculty of Science, University of Zagreb, Zagreb,  Croatia}
\author{S. N.  Zhang  }
\affiliation{Institute of Physics, Ecole Polytechnique F\'ed\'erale de Lausanne (EPFL), CH-1015 Lausanne, Switzerland}
\affiliation{National Centre for Computational Design and Discovery of Novel Materials MARVEL,
Ecole Polytechnique F\'ed\'erale de Lausanne (EPFL), CH-1015 Lausanne, Switzerland}
\author{F. Orbani\'c}
\affiliation{Department of Physics, Faculty of Science, University of Zagreb, Zagreb,  Croatia}
\author{N.  Bili\v{s}kov}
\affiliation{Ruder Bo\v{s}kovi\'c Institute,  Zagreb, Croatia }
\author{G. Eguchi}
\author{S. Paschen}
\affiliation{Institute of Solid State Physics, Vienna University of Technology, Austria }
\author{A. Kimura}
\author{X. X. Wang}
\affiliation{Department of Physical Science, Graduate School of Science, Hiroshima University, Hiroshima, Japan}
\author{T. Osada}
\author{K. Uchida}
\author{M. Sato}
\affiliation{Institute for Solid State Physics, University of Tokyo, Chiba, Japan}
\author{Q. S. Wu}
\author{O. V. Yazyev}
\affiliation{Institute of Physics, Ecole Polytechnique F\'ed\'erale de Lausanne (EPFL), CH-1015 Lausanne, Switzerland}
\affiliation{National Centre for Computational Design and Discovery of Novel Materials MARVEL,
Ecole Polytechnique F\'ed\'erale de Lausanne (EPFL), CH-1015 Lausanne, Switzerland}
\author{I. Kokanovi\'c}
\affiliation{Department of Physics, Faculty of Science, University of Zagreb,  Zagreb, Croatia}
\affiliation{Cavendish Laboratory, University of Cambridge, Cambridge, United Kingdom}

\date{\today}

\begin{abstract} 
We instigate the angle-dependent magnetoresistance (AMR) of the layered nodal-line Dirac semimetal ZrSiS for the in-plane and out-of-plane current directions.   
This material has recently revealed an intriguing  butterfly-shaped in-plane AMR that is not well understood.  
Our measurements of the polar out-of-plane AMR show a surprisingly different response with a pronounced cusp-like feature.
The maximum of the cusp-like anisotropy is reached when the magnetic field is oriented in the $a$-$b$ plane.
Moreover, the AMR for the azimuthal out-of-plane current direction exhibits a very strong four-fold $a$-$b$ plane anisotropy.
Combining the Fermi surfaces calculated from first principles with the Boltzmann's semiclassical transport theory 
we reproduce and explain all the prominent features of the unusual behavior of the in-plane and out-of-plane AMR. 
We are also able to clarify  the origin of the strong non-saturating transverse  magnetoresistance as an effect of imperfect charge-carrier compensation and open orbits.
Finally, by combining  our theoretical model and experimental data we estimate the average relaxation time of $2.6\times10^{-14}$~s and the mean free path of $15$~nm at 1.8~K in our samples of ZrSiS.
\end{abstract}


\maketitle


Square net crystal structures   have  been of a considerate interest in the structural solid-state chemistry.\cite{Tremel_1987} 
Introducing   non-trivial topology and   Dirac fermions to the field of  condensed-matter  physics has started a surge in the discovery of  new materials with linear energy dispersion.\cite{Wang_2013, Hasan_2010, Hsieh_2009, Borisenko_2014, Neupane_2014, Liu_2014, Huang_2015, Lv_2015_2, Lv_2015}
Among many square-net structures, two  phases have emerged as  especially  interesting from the topological point of view. 
The ATB$_{2}$ phase, where A stands for  alkali or rare-earth metal with +2 oxidation state, T is a 3$d$ transition metal and  B a pnictogen group  element. 
Typical representatives are Ca/Sr/Ba-MnBi$_2$ that harbor quasi-2D Dirac fermions with a highly anisotropic band dispersion in the Bi-based atomic plane along with antiferromagnetic ordering in the Mn plane.\cite{Jo_2014, Feng_2014, Li_2016, Zhang_2016, Wang_2016} 
Replacing the alkali earth metal with Eu leads to an additional inter-layer decoupling  and the formation of the half-integer quantum Hall effect.\cite{Masuda_2016}

Another interesting phase is  MX$^{'}$X$^{''}$, which incorporates a large group of compounds,\cite{Tremel_1987} where  M is a metal (Zr, Hf, Ta, Nb), X$^{'}$ is a +2 valence state of Si, Ge, As and X$^{''}$ belongs to the  chalcogen group.
The prototypical representative  of this group is ZrSiS which is the subject of this work. 
ZrSiS and isostructural compounds have recently gained  a lot of attention due to the glide and screw symmetry protected crossing of the conduction and valence bands  and thus resulting in a nodal-line Dirac semimetal (NLDSM) phase. 
 NLDSMs are the topological phases related to the 3D Dirac and Weyl semimetals with the difference being that the conduction and valence bands do not  cross only at isolated points in the $k$ space, but 
 form  loop or  nodal-line degeneracies that give rise to new and interesting physical phenomena.\cite{Rui_2018, Ramamurthy_2017,Fang_2015,  Rudenko_2018, Burkov_2018,Huh_2016}

In the absence of  spin-orbit interaction (SOI), ZrSiS has one set of nodal lines close to the  Fermi energy (E$_\mathrm{F}$) and another set located deep in the valence band.\cite{Schoop_2016, Rudenko_2018} 
It has been argued that the nodal lines located in the vicinity of  E$_\mathrm{F}$ are  protected by $C_{2v}$ symmetry and are thus  susceptible to a degeneracy lifting due to the SOI  that is effectively transforming the system into a weak topological insulator.\cite{Xu_2015} 
On the other hand,  the deep-lying nodal lines are topologically protected by the non-symmorphic symmetry. 
ZrSiS has several theoretically predicted unique properties among 3D Dirac semimetals, including
the large interval of  linear dispersion (reaching almost 2~eV) without the presence of any trivial bands and the high degree of electron-hole symmetry.\cite{Schoop_2016, Xu_2015}

Recent studies  of the Fermi surface (FS) morphology in ZrSiS by means of angle-resolved photoemission  spectroscopy (ARPES) and quantum oscillations measurements (QOM) have confirmed the nature and the position  of the two pockets: a large electron and a smaller hole pocket.\cite{Schoop_2016, Neupane_2016, Fu_2017, Hosen_2017,Matusiak_2017, Hu_2017}
The electron pocket has a 3D nature, whereas  the hole pocket  shows a quasi-2D signature.\cite{Zhang_2018, Ali_2016, Singha_2017-2} 
The QOM have also revealed the signature of another very small pocket with a puzzling Berry phase and the ultraquantum limit at around 10 T.\cite{Matusiak_2017, Hu_2017}
Due to its small size and  small charge-carrier concentration,  as well as its  weakly elongated  ellipsoidal shape observed by   QOM, this pocket is most likely irrelevant for the observed  charge transport effects under the rotation of magnetic field.\cite{Abrikosov}
Furthermore, a high-magnetic field study of ZrSiS has revealed an interesting  magnetic-breakdown effect and an unusual mass enhancement, whereas in the sister compound HfSiS  an effect of  Klein tunneling  between electron and hole pockets was detected.\cite{Pezzini_2017, Delft_2018}
The magnetoresistance is  large and unsaturated with a sub-quadratic  magnetic field  dependence as  frequently observed in the Dirac and Weyl semimetals.\cite{Singha_2017-2, Zhang_2018, Xiong-2015, Liang-2015}
On the other hand,  constant field angular-dependent magnetoresistance (AMR) measurements  have observed an unexpected and intricate butterfly-shaped anisotropy for  current  applied along  the in-plane axes.\cite{Ali_2016, Wang_2016-2, Zhang_2018, Hu_2016}  

\begin{figure}
\includegraphics[width=9cm]{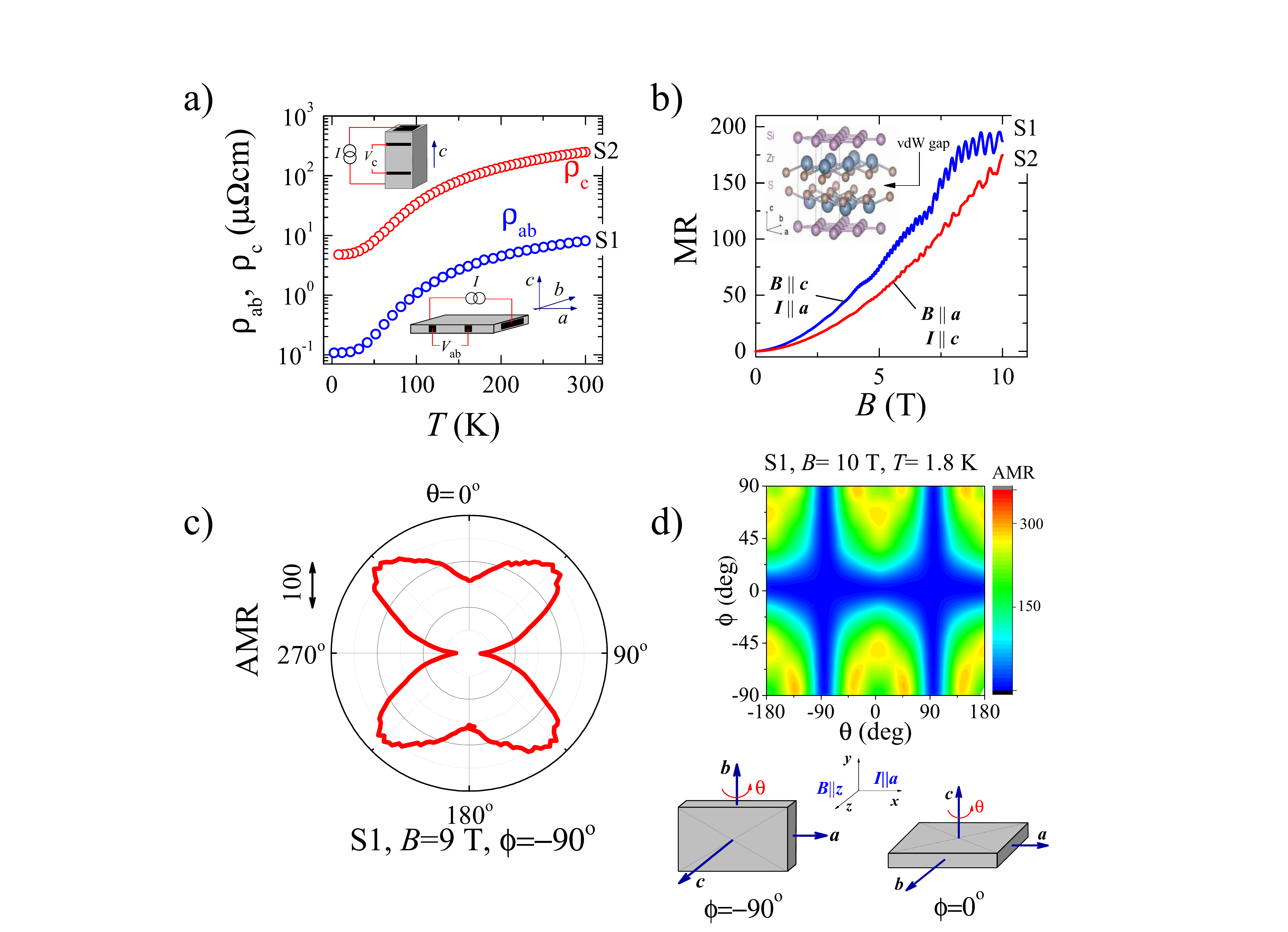}
\caption{\textbf{a)} Temperature dependent resistivity for the in-plane  ($\rho_{ab}$) and   out-of-plane  ($\rho_{c}$) current directions of the two single crystals (S1 and S2, respectively) of ZrSiS. 
In  both directions, resistivity shows metallic behavior with  moderately strong   low-$T$ anisotropy  $\rho_{c}/\rho_{ab}|_{1.8K}$ $\approx $50. 
The   low-$T$ $\rho_{ab}\approx0.1$ $\mu\Omega$cm shows excellent material quality.
\textbf{b)} For both orientations,  high-$B$ and  low-$T$ transverse magnetoresistance (MR at 1.8 K) shows a comparable magnitude, reaching almost 20 000$\%$ at $B=$10 T. 
The inset shows the tetragonal crystal structure of ZrSiS (space group $P4/nmm$). 
Adjacent layers  of S atoms are van der Waals bonded and thus form natural cleavage planes.
\textbf{c)} The intra-AMR at constant $B$ for the in-plane current direction (along a-axis) shows an unusual butterfly-like pattern for $B$  rotated in the $c$-$b$ plane.  
\textbf{d)} Detailed low-$T$ angular spectroscopic mapping  of  MR at 10~T for the in-plane current direction  provides an overview of the shape of the  AMR in 3D. 
The sketches provide  information on the rotation geometry. 
$B$ is always oriented along $z$ and the current along the $x$ axis   of the external reference frame.   
The mapping is performed by scanning   the angle $\theta$  for a fixed value of $\phi$.}
\label{fig:1}
\end{figure}

\begin{figure*}
\includegraphics[width=13cm]{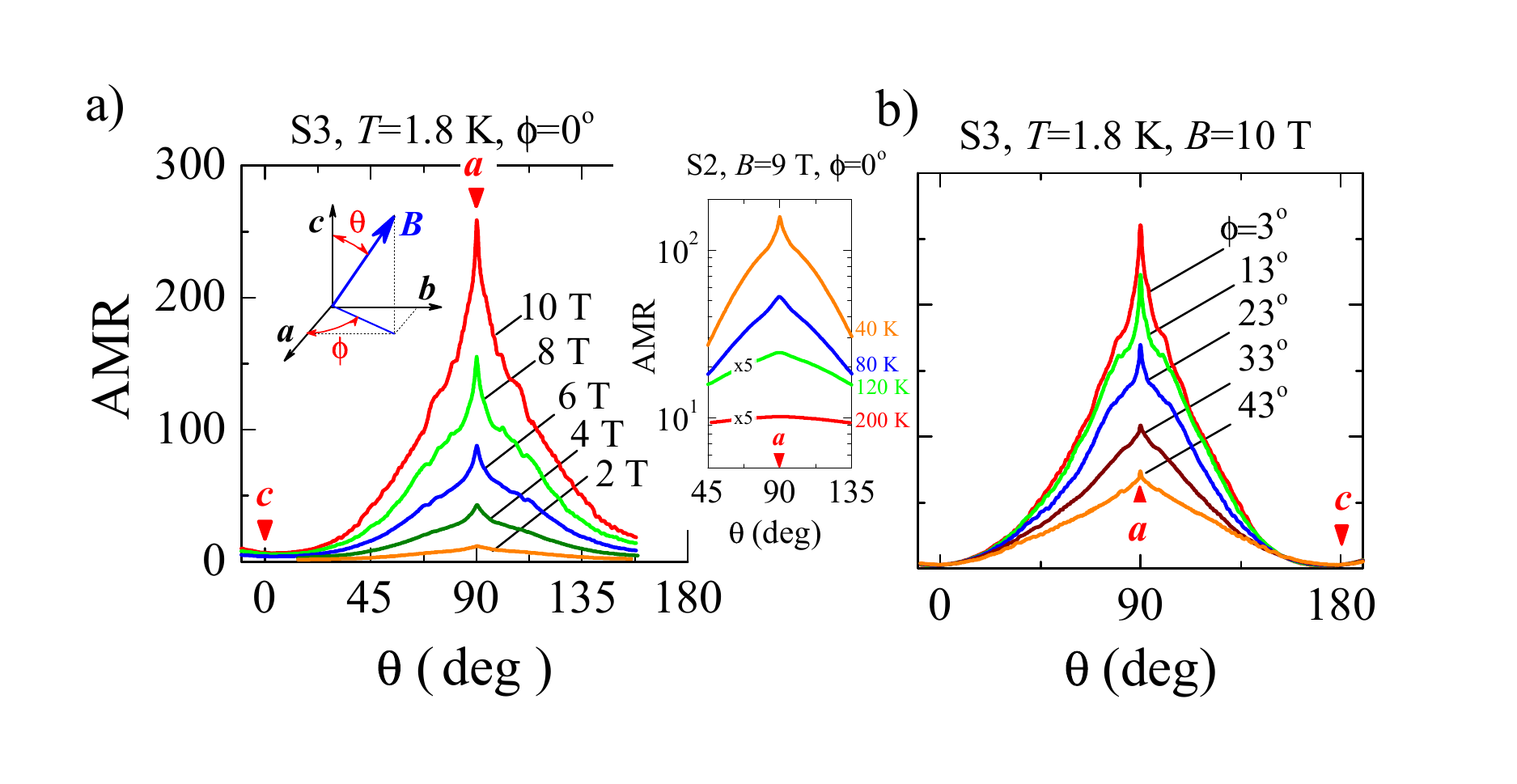}
\caption{\textbf{a)} Polar  inter-AMR  for several constant $B$ at $T=$1.8 K and $\phi=0^\circ$, with the current in  the out-of-plane direction (along the $c$-axis) for sample S3, shows strikingly different behavior  than  the intra-AMR.  $B$ is rotated in the $c$-$a$ plane.  
The AMR shows  a pronounced increase in magnitude as  $B$ is rotated  toward the $a$-$b$ plane and forms a cusp-like feature.  
The inset shows the  temperature dependence of polar inter-AMR. The anisotropy becomes very weak around 100~K.   
\textbf{b)}  Polar intra-AMR for several different directions along the azimuthal ($\phi$) angle reveals a strong $a$-$b$ plane  anisotropy.
}
\label{fig:2}
\end{figure*}

In this paper, we present a detailed study of magnetic field ($B$) and temperature ($T$) dependence  of  the AMR  in ZrSiS single crystals for  current oriented along high-symmetry directions.
The AMR has proved to be a very powerful tool for studying the FS shape of 3D and quasi-2D (q-2D) systems, with many materials exhibiting non-classical behavior.\cite{Klauder_1960, Collaudin_2015, Balicas_2005}
We have performed out-of-plane AMR measurements (inter-AMR) with  current along the $c$-direction and  in-plane AMR measurements (intra-AMR) with current along the $a$($b$)-axis.
To the best of our knowledge,    ZrSiS and related compounds have not been previously characterized  for the current oriented along the $c$ direction. 
For current  along the  $c$-direction  polar inter-AMR  reveals  a large cusp-like  anisotropy  which becomes  pronounced close to the $a$-$b$ plane.
Additionally, the azimuthal inter-AMR shows a strong $a$-$b$ plane anisotropy with four-fold symmetry and a minimum  at an angle corresponding  to an odd multiples of $\pi/4$. 
In the case of the intra-AMR, the polar scan displays a previously observed  butterfly-like shape.

To understand  the striking difference of  the AMR response for the in-plane and  out-of-plane current directions and to elucidate the role of the putative q-2D hole pocket, we have employed a theoretical transport model based on the Fermi surface calculated from first principles combined with the semi-classical Boltzmann equation. 
Using this model we were able to reproduce all features observed in  experimental data. Our model explains the intricate butterfly-like AMR and strongly anisotropic inter-AMR in terms of charge-carrier compensation due to electron and hole pockets and the effect of strong off-diagonal elements in the conductivity tensor. 
Furthermore, by using  the model we clarify the origin of large  sub-quadratic non-saturating magnetoresistance as an effect of the imperfect charge-carrier compensation. 
Finally, combining  the experimental AMR results  and  the theoretical  transport model we were able to refine  the shape of the Fermi surface and estimate the average scattering time and the mean free path.

Single crystals of ZrSiS  were grown by  chemical vapor transport  and show excellent quality with a low-$T$ in-plane resistivity of only 0.1~$\mu\Omega$cm.\cite{Ali_2016} 
Due to their layered crystal structure,  ZrSiS commonly grows in as plate-like crystals, with a thickness of around  100 $\mu$m. 
By optimization of the synthesis  procedure we  managed to obtain samples of sufficient thickness (in the mm range) which also allowed us to measure the out-of-plane transport properties.
All measured samples S1, S2, and S3 are cut from the same bigger single crystal whose homogeneity was verified before cutting.   
The zero $B$ temperature dependence of the resistivity   for the in-plane current direction $\rho_{ab}$ of sample S1 shows  metallic behavior [Fig.\ref{fig:1}a]  with a residual resistivity ratio (RRR=$\rho_{300 K}/\rho_{1.8 K}$) of around 80. 
The  out-of-plane resistivity $\rho_{c}$ of sample S2 shows also  metallic behavior  but with  a considerably higher resistivity contributing to a moderate anisotropy $\rho_{c}/\rho_{ab}$ of around  50 at the lowest measured temperature.
The almost identical $T$ profile for the in-plane and out-of-plane transport points towards the  coherent intra-layer transport, which is expected in transport dominated by a 3D-FS. 

The transverse magnetoresistance (MR)\cite{foot003} given in Fig.~\ref{fig:1}b shows a very strong response reaching almost 20 000~$\%$ at 10 T for both orientations (S1: $B||c$, $I || a$, and S2: $B||a$, $I||c$). 
A strong, non-saturating MR is a commonly  observed property  of  3D Dirac and Weyl semimetals arising from   multiband transport of high mobility   charge carriers. 
The MR of samples S1 and S2 has a sub-quadratic $B$ dependence\cite{foot002} which has been recently associated with a $B$ dependent  mobility.\cite{Fauque_2018, Zhao_2018}
Sample S1 displays strong  Shubnikov-de Haas (SdH) quantum oscillations with frequencies corresponding to 8.5~T and 241~T.
ARPES measurements unambiguously related the higher frequency (241 T) to the q-2D tube-like hole pocket, whereas the position and exact shape of the pocket  with the smaller frequency is not yet determined. 
Due to its small size and  weakly elongated  ellipsoidal shape   its contribution to the total  magnetotransport of the charge carriers  should be negligible.
SdH oscillations observed in sample S2 are less pronounced and composed of several frequencies (17~T, 23~T and 170~T). 

Under  the polar rotation (angle $\theta$) of sample S1 at $B=9$~T the transverse intra-AMR,\cite{foot003} shown in Fig.~\ref{fig:1}c) exhibits a peculiar four-fold  butterfly-shaped angular dependence with the angle of  maximum resistivity at odd multiples of $\theta=\pi/4$. 
The origin of this peculiar intra-AMR has been elusive since previously this kind of AMR was only observed in magnetically-ordered materials where its origin is not purely orbital.\cite{Jovan_2010}     
By using the framework of the semiclassical transport model we were able to  understand the origin of the intra-AMR in terms of the charge-carrier compensation effects of the electron and hole pockets. 
Figure~\ref{fig:1}d displays a detailed spectroscopic mapping of the intra-AMR  transport.  
When $B$ is tilted away from the $c$-axis  the AMR increases showing the butterfly-shaped profile for all values of the azimuthal angle $\phi$. 
On the other hand,  for the in-plane rotation ($a$-$b$ plane) the intra-AMR is small and the anisotropy is  weak.

\begin{figure}
\includegraphics[width=9cm]{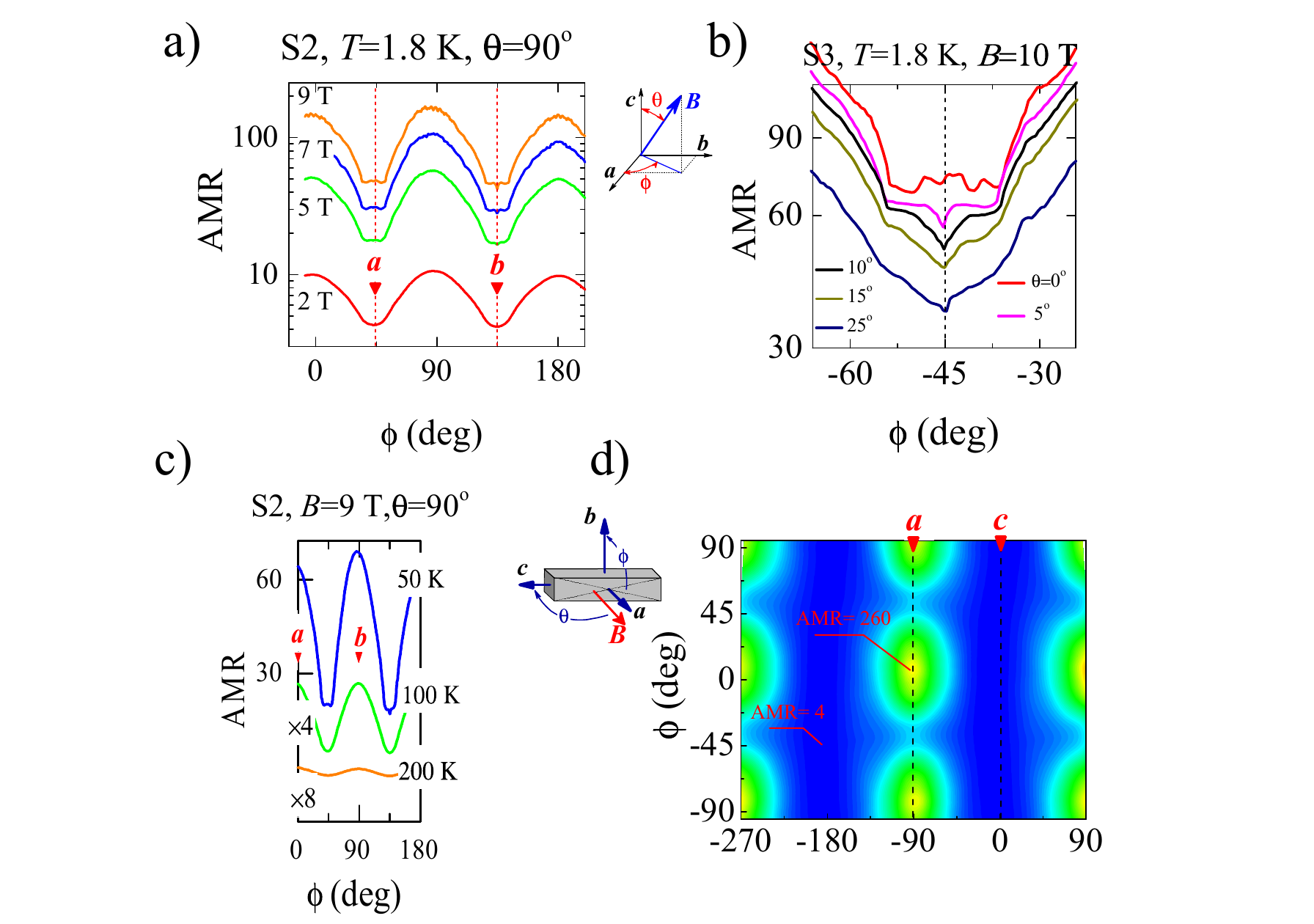}
\caption{\textbf{a)} Azimuthal  inter-AMR ($a$-$b$ plane field rotation) for sample S2 at several different $B$. 
The azimuthal AMR  shows a strong four-fold  in-plane anisotropy with the  maxima for $B$ oriented along the high-symmetry in-plane axes and the  minima at the bisector axes (odd multiples of $\phi=\pi/4$).  
At around 5~T,   in the vicinity of the bisector  axis, the AMR becomes truncated (flat) in contrast to the 2~T scan which has a quadratic-like shape.  
For $B$ close to the  high-symmetry axes, quantum oscillations can be observed. 
\textbf{ b)} It is interesting to notice (sample S3 at 10~T) that the truncation  of the inter-AMR has a substructure with a dip when  $B$ is slightly misaligned with respect to the $a$-$b$ plane   and a more complicated structure for  $B$ in the $a$-$b$ plane. The substructure can not be due to quantum oscillations, since the peak positions do not change with the field. 
The substructure probably originates from a peculiarity  in the  FS shape.   
\textbf{c)} Temperature dependence of the azimuthal inter-AMR for sample S2. Above  100~K the AMR becomes  weakly anisotropic and at round 200~K it starts  following the classical sinusoidal behavior.
\textbf{d)} Details of wide-angle spectroscopic mapping of the inter-AMR  of sample S3 at 10~T. The sketch depicts the measurement geometry. }
\label{fig:3}
\end{figure}

Figure \ref{fig:2}a  presents details of the polar scan of the inter-AMR at several discrete values of $B$ between 2~T and 10~T  for  current along  the $c$-axis for sample S3. 
For the longitudinal configuration ($B$ and current are along the $c$-axis) the MR is small for all measured $B$, which  is expected due to the vanishing Lorentz force. 
In the  model of a single spherical FS it should be zero. 
By tilting $B$  away from the $c$-axis  (in the $c$-$a$ plane)  the AMR shows a strong increase in  magnitude that becomes more pronounced with increasing  $B$ and forms a cusp-like feature.   
The   anisotropy ratio of  the MR for the longitudinal and transverse $B$ orientations is around 50 at 10~T. 
The strong increase of inter-AMR when $B$ is close to the in-plane direction is a feature commonly observed in  q-2D materials and  it is  associated with  coherent intra-layer transport, i.e. a small warping of the 2D Fermi surface.\cite{Hanasaki_1998, McKenzie_1998} 
The coherence peak is usually  accompanied by Yamaji oscillations in inter-AMR, but is not detected in our samples.\cite{Yagi_1990}
The small oscillations in the AMR profile at higher $B$  originate from the quantum oscillations, which is supported by the field dependence of the oscillations peak positions. 
Recently, in several Dirac semimetals with  square-net structure,  the peak-like response has been observed for $B$ close to the $a$-$b$ plane, which was explained  by  q-2D FS\cite{Jo_2014, Wang_2016} or by the inter-layer quantum transport governed by the Dirac point.\cite{Liu-2017}
The inset in Fig.\ref{fig:2}a shows temperature dependence of the inter-AMR  for sample S2 at 9~T.
By increasing  $T$ the cusp-like shape of the inter-AMR weakens  and, at around 200~K, it acquires the classical sinusoidal shape. 
Performing the same polar scans but now for different angles $\phi$,  we observe an indication of a strong  in-plane anisotropy  [see Fig.~\ref{fig:2}b]. 
The  polar inter-AMR  becomes significantly  weaker as $\phi$ approaches  $\pi/4$.

\begin{figure*}
\includegraphics[width=15cm]{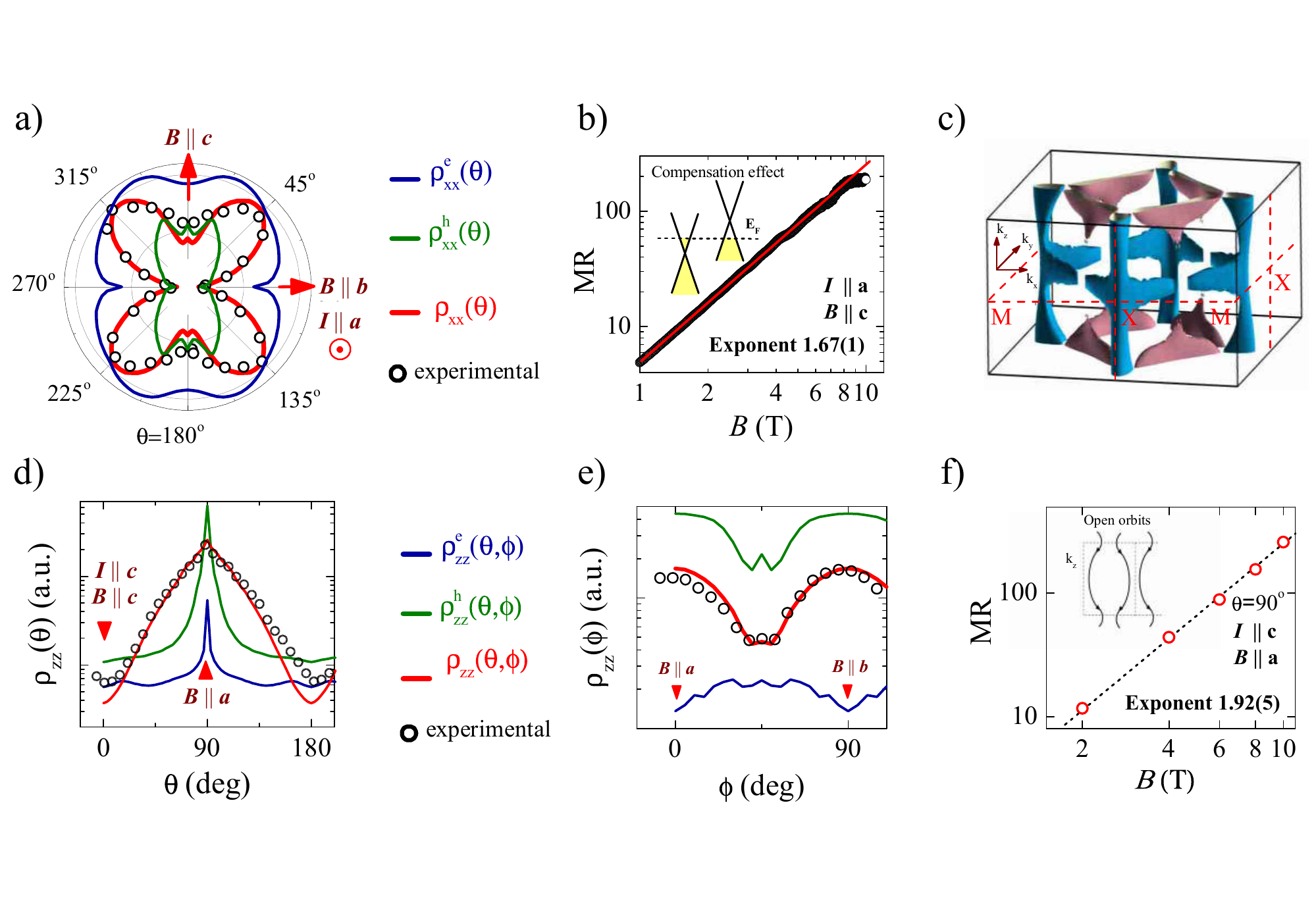}
\caption{\textbf{a)} Calculated angular resistivity $\rho_{xx}(\theta)$ (red solid line) at constant $B$ for the intra-layer charge transport. The current and magnetic field orientations are indicted. The calculations reproduce the two-fold symmetry butterfly-shaped magnetoresistance in good agreement with experimental data (black open circles). The number of experimental points has been reduced for clarity. Green and blue solid lines represent diagonal contributions of the hole ($\rho^h_{xx}$) and  electron ($\rho^e_{xx}$) pocket individually. 
\textbf{b)} The calculated MR for magnetic field along the $c$-axis ($I || a$) follows the $B^{1.68}$ dependence, which agrees perfectly with the measured exponent of $1.67(1)$, pointing to the charge-carrier compensation mechanism of the observed large transverse MR. 
\textbf{c)} Fermi surface used for calculating the conductivity tensor. The FS consist of four isolated electron pockets at $k_{z}=\pi/c$ and four hole pockets. The hole pocket hosts open orbits and has protruding arm-like features in the $k_{z}=0$ plane.
\textbf{d),e)}  The model (red line) agrees well with experimental data (open circles) for polar and azimuthal experimental inter-AMR reproducing all  distinctive  features. Green ($\rho^h_{zz}$) and  blue ($\rho^e_{zz}$) lines depict diagonal matrix elements of the resistivity for the hole and electron pockets, respectively. 
\textbf{f)}  MR values at the cusp maxima for the polar inter-AMR  reproduced  from  Fig.~\ref{fig:2}a for different values of $B$. The slope corresponds to $1.92(5)$, which is close to 1.98 predicted by our model. 
	}
\label{fig:4}
\end{figure*}

The azimuthal  inter-AMR ($\phi$ angle rotation)  given in Fig.~\ref{fig:3}a shows a strong fourfold $a$-$b$ plane anisotropy  with the maxima positioned along the high-symmetry axes $a$ and $b$ and minima along the bisector axes (odd multiples of $\pi/4$). 
The strength of the anisotropy (ratio of the  maximum and minimum values) is almost $B$ independent in the measured range between 5~T and 9~T and its value is around 4. This is in  contrast to the anisotropy of the  polar inter-AMR that is continuously growing with $B$.
The observed $a$-$b$ plane  anisotropy is fairly large. For comparison, it is roughly a factor of 2 larger than that in Sr$_2$RuO$_4$.\cite{Omichi_2000, Balicas_2005}
For larger $B$ ($>5$~T), the AMR becomes truncated close to the bisector axis, whereas at  2~T the truncation is not observed. 
The effect of  quantum oscillations is clearly seen at 9~T  close to the high-symmetry axes. 
Closer inspection  reveals that the truncation has a complicated structure, details for sample S3  at 10~T are shown in Fig.~\ref{fig:3}b. 
When $B$ is slightly misaligned with respect to the in-plane orientation, the dip in the AMR appears at the  bisector axis,  whereas when $B$ is in-plane, the truncated part shows oscillating-like behavior which is $B$ independent and thus cannot be related to quantum oscillations. 
This unusual behavior is probably  related to the local morphology  of the FS.  
Considering the temperature dependence of the azimuthal inter-AMR it can be seen that, above 100~K, the truncated part disappears [see Fig.~\ref{fig:3}c] and  the anisotropy of the  in-plane  AMR weakens.  
 In Fig.~\ref{fig:3}d we present a detailed spectroscopic mapping of the inter-AMR. It can be seen that the AMR pattern has  a twofold symmetry for $\theta$-angle rotation and fourfold symmetry for $\phi$-angle rotation. The maximum in the AMR only appears  when $B$ is oriented along the high-symmetry in-plane directions.


In order to achieve a deeper understanding of the unusual AMR response in ZrSiS we have employed numerical modeling using the semiclassical Boltzmann transport theory\cite{Liu_2009,ShengNan_2018} as implemented in WannierTools open-source software package.\cite{WannierTools} A detailed description of the methodology along with representative examples is given in Ref.~\onlinecite{ShengNan_2018}. 
We have used Fermi surfaces calculated using DFT to obtain the Boltzmann conductivity tensor $\sigma_{ij}$ for electron and hole pockets which was then transformed into the total resistivity tensor $[\rho_{ij}]=[\sigma^e_{ij}+\sigma^h_{ij}]^{-1}$. Analyzing individual contributions of electron and hole pockets will help us clarifying the physical mechanism underlying the discussed megnetotransport properties. 

As a first step, in Fig.~\ref{fig:4}a we present the transverse angular resistivity $\rho_{xx}(\theta)$ calculated at a fixed magnitude of $B$ for the intra-layer current direction. The calculations reproduce very well the experimental data (black open circles), and in particular the two-fold symmetry butterfly-like shape.
The green and blue lines in Fig.~\ref{fig:4}a represent individual contributions of the electron ($\rho^e_{xx}$) and hole ($\rho^h_{xx}$) pockets, respectively.
Closer inspection reveals that the total angular resistivity is not a simple sum of electron and hole (parallel) channels. 
Only for the high-symmetry directions, i.e. $B || c$ and $B || b$, the total calculated resistivity  is smaller that the resistivity of the contributing channels, giving rise to a noticeable drop in the intra-AMR. 
For other directions we cannot apply  the parallel channel rule due to a significant contribution of the off-diagonal elements of the conductivity tensor, from which $\rho_{xx}(\theta)$ is obtained. 

Next, we use our model to understand the mechanism of the observed large non-saturating  transverse MR  for the in-plane current direction.
While non-saturating MR with $B^2$ dependence is usually assigned to systems with perfect electron-hole compensation, deviations from the ideal quadratic scaling often observed in the high mobility systems with electron and hole pockets has recently been attributed to the field-dependent mobility.\cite{Fauque_2018}  
The transverse MR for $B || c$ scales as $B^{1.68}$, which is in excellent agreement with the measured $B^{1.67(1)}$ dependence [Fig.~\ref{fig:4}b]. 
The strong orbital MR in our case comes from the imperfect compensation effect between electron and hole pockets.  

We now proceed with the analysis  of inter-AMR with current oriented along the $c$-direction for two distinct measurement configurations -- polar ($\theta$-scan) and azimuthal ($\phi$-scan) ones.   
Fig.~\ref{fig:4}d shows the calculated polar angular resistivity $\rho_{zz}(\theta)$ (red solid line) for the current and field orientations defined in the figure.
The calculated resistivity  has a strong angular dependence with a cusp-like shape  and a maximum being reached for the in-plane oriented field. 
Comparison of the calculated and measured (black solid circles) angular dependences again shows good agreement.   
To understand the origin of this unusual behavior  we examine  the individual contributions of the electron and hole pockets, $\rho^{e}_{zz}$ and $\rho^{h}_{zz}$, respectively.
Comparison of the calculated magnetoresistivity $\rho_{zz}(\theta)$ with individual components  again shows that the total resistivity cannot be described by combining two  parallel channels and the  off-diagonal elements play a significant contribution.  
 Both  matrix elements have a strong peak for $B$ close to the in-plane orientation, but of distinct origins. 
The peak in  $\rho^{h}_{zz}$ originates from the open orbits, whereas the peak in $\rho^{e}_{zz}$ comes from the electron pocket deviates strongly from the ideal free-electron spherical shape, i.e. it results from its flatness.  

As a next step, we aim at understanding  the $a$-$b$ plane anisotropy of the intra-AMR [Fig.~\ref{fig:4}e].
Good agreement of the calculated and experimental angular dependences is achieved as in the previous cases. 
Comparing  $\rho_{zz}(\phi)$ with the contributions of individual pockets $\rho^{h}_{zz}(\phi)$ (green line) and $\rho^{e}_{zz}(\phi)$ (blue line),
allows us to conclude that the   angular dependence of $\rho_{zz}(\phi)$ is mostly due to the hole pocket.
The truncated part of the azimuthal AMR  most likely also has its origin in the hole pocket since $\rho^{h}_{zz}$ shows anomalous behavior close to $\pi/4$. 
Furthermore, in Fig.~\ref{fig:4}f we have traced the cups maxima [from Fig.~\ref{fig:2}a] as a function of $B$ obtaining the $B^{1.92(5)}$ scaling.
Our calculations predict a comparable value of 1.98 and both results are very close to the quadratic behavior expected for the open-orbit mechanism. 
Even though we have observed almost quadratic dependence, the azimuthal in-plane anisotropy  indicates that besides the dominant effect of open orbits, the charge-carrier compensation effect also plays an important role since open orbits cannot  account  for the observed azimuthal  angular dependence. 
Recently, strong increase in  the inter-AMR close to the $a$-$b$ plane has been  reported in several similar  square-net Dirac materials.\cite{ Jo_2014, Wang_2016}    
It has been argued  that in these materials  the peak-like structure  originates form the small closed orbits on the side of the  corrugated q-2D FS,\cite{Ghannadzadeh_2017, Hanasaki_1998}  
whereas in our case strong increase in the  AMR is  mostly due to  the open orbits.  

Combining the measured data with the theoretical model  we are now able to comment on the average relaxation time $\tau$ and the FS shape. 
By using the constant relaxation time approximation and fitting the semiclassical Boltzmann model to the Hall measurements\cite{Sup} we obtained $\tau\approx 2.6 \times10^{-14}$~s. 
From the ARPES measurements, the  Fermi velocity  is  estimated to be around  $5-6 \times10^{5}$~m/s  resulting in the mean free path  $l=15$~nm.\cite{Sup} 
Assuming that at 2~K impurity scattering is the dominant  charge-carrier relaxation mechanism we can estimate impurity concentration of ca. $10^{17}$ cm$^{-3}$.
Recent publication on ZrSiS proposed several different FS morphologies.\cite{Fu_2017, Pezzini_2017, Ali_2016}
These discrepancies can be related to the sensitivity of the FS shape calculated using DFT to the exact position of  $E_\mathrm{F}$ and various details of the methodology, such as pseudopotentials used in the calculations.  
The reported FS shapes have similar structure of the electron and hole pockets in the $k_z=\pi/c$ plane,  but differ significantly in the $k_z=0$ plane. 
Fig.~\ref{fig:4}c shows our calculated FS characterized by the hole pocket consisting of an elongated tube-like structure that gives rise to  open orbits. 
In the $k_z=0$ plane the hole pocket has protruding  "arm-like" features extending along the X-X lines.
Good agreement between all the peculiar features of the calculated and experimental AMR provides a strong indication that the calculated FS reproduces the real one.

In conclusion, we have presented a detailed study of the angular magnetoresistance (AMR) in the nodal-line Dirac semimetal ZrSiS. 
We have determined the low-temperature, zero-field  anisotropy between the in-plane and out-of-plane directions to be moderately strong with typical values around 50. 
 The AMR was measured in two configurations, for current oriented in-plane along the $a$-axis (intra-AMR) and out-of-plane  along the $c$-axis (inter-AMR). 
The intra-AMR shows an unusual butterfly-like shape previously reported by other authors. 
The inter-AMR shows a strong cusp-like shape anisotropy for  polar angle rotation, with a maximum achieved for the magnetic field oriented in the $a$-$b$ plane.
Additionally,  the azimuthal angle rotation shows strong anisotropy with a  four-fold symmetry, with the minimum at odd multiples of  $\phi=\pi/4$.
In order to understand this intricate  AMR we have employed  a theoretical model based on the Fermi surface calculated from first principles and the  Boltzmann semiclassical theory.  The model successfully reproduced all observed features of both the inter- and intra-AMR. 
Furthermore, our model explains the sub-quadratic dependence of transverse magnetoresistance as an effect of imperfect charge-carrier compensation for the in-plane case, and an open-orbit mechanism combined with charge-carrier compensation for the out-of-plane current direction. 
We were able to estimate the average relaxation time to be  around $\tau\approx 2.6 \times10^{-14}$ s, the mean free path $l\approx15$ nm and  more accurately determine the Fermi surface shape.

\begin{acknowledgments}
This work has been supported in part by the Croatian Science Foundation under the project (IP-2018-01-8912).  
N.M. thanks the ISSP, University of Tokyo for  financial support. 
S.N.Z., Q.S.W., O.V.Y. acknowledge support by the NCCR Marvel. M.N, I.K. and F.O. acknowledge the support of project CeNIKS co-financed by the Croatian Government and the European Union through the European Regional Development Fund - Competitiveness and Cohesion Operational Programme (Grant No. KK.01.1.1.02.0013). We acknowledge help of A. Dra\v{s}ner in the material  synthesis. First-principles and transport calculations have been performed at the Swiss National Supercomputing Centre (CSCS) under Project No. s832 and the facilities of Scientific IT and Application Support Center of EPFL. G.E. and S.P. acknowledge the Austrian Science Fund (project I2535-N27). N.B. acknowledge support from the Croatian Science Foundation (project PKP-2016-06-4480).

 \end{acknowledgments}

\bibliography{library}

\begin{thebibliography}{50} 
\bibitem{Tremel_1987}
W. Tremel, and R. Hoffmann,   J. Am. Chem. Soc.109, 124 (1987). 
\bibitem{Hasan_2010}
M. Z. Hasan,  and C. K.  Kane,   Rev. Mod. Phys. 82, 3045 (2010).
\bibitem{Hsieh_2009}
D. Hsieh, Y. Xia, D. Qian, L. Wray, F. Meier, J. H. Dil, J. Osterwalder, L. Patthey, A. V. Fedorov, H. Lin, A. Bansil, D. Grauer, Y. S. Hor, R. J. Cava, and M. Z. Hasan, Phys. Rev. Lett. 103, 146401 (2009).
\bibitem{Wang_2013}
Z. Wang, H. Weng, Q. Wu, X. Dai, and Z. Fang,  Phys. Rev. B 88, 125427 (2013).
\bibitem{Borisenko_2014}
S. Borisenko, Q. Gibson, D. Evtushinsky, V. Zabolotnyy, B. Buchner, and R. J. Cava, 
Phys. Rev. Lett. 113, 027603  (2014).
\bibitem{Neupane_2014}
M. Neupane, S. Xu, R. Sankar, N. Alidoust, G. Bian, C. Liu, I. Belopolski, T.-R. Chang, H.-T. Jeng, H. Lin, A. Bansil, F. Chou, and M.Z. Hasan, Nat. Commun. 5, 3786 (2014).
\bibitem{Liu_2014}
Z. K. Liu \textit{et}. al., Nat. Mater.  13,  677 (2014)
\bibitem{Lv_2015}
B.Q. Lv, H.M. Weng, B.B. Fu, X.P. Wang, H. Miao, J. Ma, P. Richard, X.C. Huang, L.X. Zhao, G.F. Chen, Z. Fang, X. Dai, T. Qian, and H. Ding, Phys. Rev. X 5, 031013 (2015).
\bibitem{Lv_2015_2}
B.Q. Lv, N. Xu, H.M. Weng, J.Z. Ma, P. Richard, X.C. Huang, L.X. Zhao, G.F. Chen, C.E. Matt, F. Bisti, V.N. Strocov, J. Mesot, Z. Fang, X. Dai, T. Qian, M. Shi, and H. Ding, Nat. Phys. 11, 724 (2015).
\bibitem{Huang_2015}
S.M. Huang, S.Y. Xu, I. Belopolski, C.C. Lee, G. Chang, B. Wang, N. Alidoust, G. Bian, M. Neupane, C. Zhang, S. Jia, A. Bansil, H. Lin, and M.Z. Hasan, Nat. Commun. 6, 1 (2015).
\bibitem{Jo_2014}
Y.J. Jo, J. Park, G. Lee, M.J. Eom, E.S. Choi, J.H. Shim, W. Kang, and J.S. Kim, Phys. Rev. Lett. 113, 1 (2014).
\bibitem{Feng_2014} 
 Y. Feng, Z. Wang, C. Chen, Y. Shi, Z. Xie, H. Yi, A. Liang, S. He, J. He, Y. Peng, X. Liu, Y. Liu, L. Zhao, G. Liu, X. Dong, J. Zhang, C. Chen, Z. Xu, X. Dai, Z. Fang, and X.J. Zhou, Sci. Rep. 4, 1 (2014).
\bibitem{Li_2016} 
L. Li, K. Wang, D. Graf, L. Wang, A. Wang, and C. Petrovi\'c, Phys. Rev. B 93, 1 (2016).
\bibitem{Zhang_2016}
A. Zhang, C. Liu, C. Yi, G. Zhao, T. Xia, J. Ji, Y. Shi, R. Yu, X. Wang, C. Chen, and Q. Zhang, Nat. Commun. 7, 13833 (2016).
\bibitem{Wang_2016}
A. Wang, D. Graf, L. Wu, K. Wang, E. Bozin, Y. Zhu, and C. Petrovic, Phys. Rev. B 94, 125118 (2016).
\bibitem{Masuda_2016}
H. Masuda, H. Sakai, M. Tokunaga, Y. Yamasaki, A. Miyake, J. Shiogai, S. Nakamura, S. Awaji, A. Tsukazaki, H. Nakao, Y. Murakami, T. -h. Arima, Y. Tokura, and S. Ishiwata, Sci. Adv. 2, e1501117 (2016).
\bibitem{Ramamurthy_2017}
S.T. Ramamurthy and T.L. Hughes, Phys. Rev. B 95, 075138 (2017).
\bibitem{Fang_2015}
C. Fang, Y. Chen, H.-Y. Kee, and L. Fu, Phys. Rev. B 92, 081201 (2015).
\bibitem{Rudenko_2018}
A.N. Rudenko, E.A. Stepanov, A.I. Lichtenstein, and M.I. Katsnelson, Phys. Rev. Lett. 120, 216401 (2018).
\bibitem{Burkov_2018}
A.A. Burkov, Phys. Rev. B 97, 165104 (2018).
\bibitem{Rui_2018}
W.B. Rui, Y.X. Zhao, and A.P. Schnyder, Phys. Rev. B 97, 161113 (2018).
\bibitem{Huh_2016}
Y. Huh, E.-G. Moon, and Y.B. Kim, Phys. Rev. B 93, 035138 (2016).
\bibitem{Schoop_2016}
L.M. Schoop, M.N. Ali, C. Straßer, A. Topp, A. Varykhalov, D. Marchenko, V. Duppel, S.S.P. Parkin, B. V. Lotsch, and C.R. Ast, Nat. Commun. 7, 11696 (2016).
\bibitem{Xu_2015}
Q. Xu, Z. Song, S. Nie, H. Weng, Z. Fang, and X. Dai, Phys. Rev. B 92, 205310 (2015).
\bibitem{Neupane_2016}
M. Neupane, I. Belopolski, M.M. Hosen, D.S. Sanchez, R. Sankar, M. Szlawska, S.-Y. Xu, K. Dimitri, N. Dhakal, P. Maldonado, P.M. Oppeneer, D. Kaczorowski, F. Chou, M.Z. Hasan, and T. Durakiewicz, Phys. Rev. B 93, 201104 (2016).
\bibitem{Fu_2017}
B. Fu, C. Yi, T. Zhang, M. Caputo, J. Ma, X. Gao, B. Lv, L. Kong, Y. Huang, M. Shi, S. Vladimir, C. Fang, H. Weng, Y. Shi, T. Qian, and H. Ding, ArXiv 1712.00782 (2017). 
\bibitem{Hosen_2017}
M.M. Hosen, K. Dimitri, I. Belopolski, P. Maldonado, R. Sankar, N. Dhakal, G. Dhakal, T. Cole, P.M. Oppeneer, D. Kaczorowski, F. Chou, M.Z. Hasan, T. Durakiewicz, and M. Neupane, Phys. Rev. B 95, 161101 (2017).
\bibitem{Matusiak_2017}
M. Matusiak, J.R. Cooper, and D. Kaczorowski, Nat. Commun. 8, 15219 (2017).
\bibitem{Hu_2017}
J. Hu, Z. Tang, J. Liu, Y. Zhu, J. Wei, and Z. Mao, Phys. Rev. B 96, 045127 (2017).
\bibitem{Zhang_2018}
J. Zhang, M. Gao, J. Zhang, X. Wang, X. Zhang, M. Zhang, W. Niu, R. Zhang, and Y. Xu, Front. Phys. 13, 1 (2018).
\bibitem{Singha_2017-2}
R. Sankar, G. Peramaiyan, I.P. Muthuselvam, C.J. Butler, K. Dimitri, M. Neupane, G.N. Rao, M.-T. Lin, and F.C. Chou, Sci. Rep. 7, 40603 (2017).
\bibitem{Ali_2016}
M.N. Ali, L.M. Schoop, C. Garg, J.M. Lippmann, E. Lara, B. Lotsch, and S.S.P. Parkin, Sci. Adv. 2, e1601742 (2016).
\bibitem{Abrikosov}
A. A. Abrikosov, \textit{Fundamentals of the Theory of Metals}, Reprint of the North Holland, Amsterdam, 1988.
\bibitem{Pezzini_2017}
 S. Pezzini, M.R. van Delft, L.M. Schoop, B. V. Lotsch, A. Carrington, M.I. Katsnelson, N.E. Hussey, and S. Wiedmann, Nat. Phys. 14, 178 (2017).
\bibitem{Delft_2018}
M.R. van Delft, S. Pezzini, T. Khouri, C.S.A. Mueller, M. Breitkreiz, L.M. Schoop, A. Carrington, N.E. Hussey, and S. Wiedmann, Phys. Rev. Lett. 121, 256602 (2018).









\bibitem{Xiong-2015}
J. Xiong, S. K. Kushwaha, T. Liang, J. W. Krizan, M. Hirschberger, W. Wang, R. Cava, N. P. Ong,Science 350, 413–416 (2015).
\bibitem{Liang-2015}
T. Liang, Q. Gibson, M. N. Ali, M. Liu, R. J. Cava, N. P. Ong,  Nat. Mater. 14, 280–284 (2015).


\bibitem{Wang_2016-2}
 X. Wang, X. Pan, M. Gao, J. Yu, J. Jiang, J. Zhang, H. Zuo, M. Zhang, Z. Wei, W. Niu, Z. Xia, X. Wan, Y. Chen, F. Song, Y. Xu, B. Wang, G. Wang, and R. Zhang, Adv. Electron. Mater. 2, 1600228 (2016).
\bibitem{Hu_2016}
Jin Hu, Zhijie Tang, Jinyu Liu, Xue Liu, Yanglin Zhu, David Graf, Yanmeng Shi, Shi Che, Chun Ning Lau, Jiang Wei, and Zhiqiang Mao,
Phys. Rev. Lett. \textbf{117}, 016602 (2016). 
\bibitem{Klauder_1960}
J. R. Klauder and J. E. Kunzler, \textit{The Fermi Surface} (Wiley, New York, 1960).
\bibitem{Collaudin_2015}
A. Collaudin, B. Fauqué, Y. Fuseya, W. Kang, and K. Behnia, Phys. Rev. X 5, 021022  (2015).
\bibitem{Balicas_2005}
L. Balicas, S. Nakatsuji, D. Hall, T. Ohnishi, Z. Fisk, Y. Maeno, and D.J. Singh, Phys. Rev. Lett. 95, 196407 (2005).
\bibitem{foot003} 
The experimental magnetoresistance is defined as MR$=[\rho(B)-\rho_0]/\rho_0$, whilst the experimental polar and azimuthal AMR  at a constant $B$ are defined as  AMR$=[\rho(\theta)-\rho_0]/\rho_0$, AMR$=[\rho(\phi)-\rho_0]/\rho_0$, respectively,  $\rho_0$ is the zero field value.
\bibitem{foot002} For sample S1 $\rho_{ab}(B) \propto B^{1.68(1)}$,  and for sample S2  $\rho_{c}(B) \propto B^{1.82(3)}$, which deviates from the theoretically predicted value of the exponent 1.98 due to misalignment of $B$ with respect to the high-symmetry axis.
\bibitem{Zhao_2018}
L. Zhao, L. Xu, H. Zuo, X. Wu, G. Gao, and Z. Zhu,   Phys. Rev. B  98, 085137 (2018).
\bibitem{Fauque_2018}
Benoît Fauqué, Xiaojun Yang, Wojciech Tabis, Mingsong Shen, Zengwei Zhu, Cyril Proust, Yuki Fuseya, and Kamran Behnia
Phys. Rev. Materials 2, 114201  (2018).
\bibitem{Jovan_2010}
V. Jovanovi\'c, L. Fruchter, Z. Li, and H. Raffy, Physical Review B 81, 134520 (2010).
\bibitem{Hanasaki_1998}
N. Hanasaki, S. Kagoshima, T. Hasegawa, T. Osada, and N. Miura, Phys. Rev. B 57, 1336 (1998).
\bibitem{McKenzie_1998}
R.H. McKenzie and P. Moses, Phys. Rev. Lett. 81, 4492 (1998).
\bibitem{Yagi_1990}
R. Yagi, Y. Iye, T. Osada, and S. Kagoshima, J. Phys. Soc. Japan 59, 3069 (1990).
\bibitem{Liu-2017}
J.Y. Liu, J. Hu, D. Graf, T. Zou, M. Zhu, Y. Shi, S. Che, S.M.A. Radmanesh, C.N. Lau, L. Spinu, H.B. Cao, X. Ke, and Z.Q. Mao, Nat. Commun. 8, 646 (2017).
\bibitem{Omichi_2000}
E. Ohmichi, Y. Maeno, S. Nagai, Z.Q. Mao, M.A. Tanatar, and T. Ishiguro, Phys. Rev. B 61, 7101 (2000).
%
\bibitem{Liu_2009}
Y. Liu, H.-J. Zhang, and Y. Yao, Phys. Rev. B  79, 245123 (2009).
\bibitem{ShengNan_2018} 
ShengNan Zhang, QuanSheng Wu, Yi Liu, and Oleg V. Yazyev,  Phys. Rev. B 99, 035142 (2019).
%
\bibitem{WannierTools}
Q. Wu, S. Zhang, H.-F. Song, M. Troyer, and A. A. Soluyanov, Comput. Phys. Commun. 224, 405 (2018).
\bibitem{Ghannadzadeh_2017}
S. Ghannadzadeh, S. Licciardello, S. Arsenijevi\'c, P. Robinson, H. Takatsu, M.I. Katsnelson, and N.E. 
Hussey, Nat. Commun. 8, 15001 (2017). 
\bibitem{Sup}
See the Supplementary data.










\end{thebibliography}

\end{document}